\documentclass[journal]{IEEEtran}

\usepackage{cite}

\ifCLASSINFOpdf
\else
\fi
\usepackage[cmex10]{amsmath}

\usepackage{array}
\usepackage{graphicx}
\usepackage{caption}
\usepackage{subcaption}
\usepackage{color}

\usepackage{stfloats}

\usepackage{color,soul}

\usepackage{amsmath,amsfonts,amssymb,amscd}
\usepackage{mathrsfs}
\usepackage{graphicx,epsfig}

\usepackage{cases}
\usepackage{multirow}

\usepackage[font=singlespacing]{caption}
\usepackage{lscape}
\usepackage{verbatim}
\usepackage{epstopdf}
\usepackage{balance}

\usepackage{psfrag}
\psfrag{J}{$\mathbb{E}$}

\usepackage{bbm}

\usepackage{lipsum}

\usepackage{mathrsfs}
\usepackage{balance}
\usepackage{soul}
\usepackage{graphicx}

\usepackage{caption}
\usepackage{subcaption}

\usepackage{xcolor}

\begin{document}

\title{Deep Multi-Agent Reinforcement Learning for Cost Efficient Distributed Load Frequency Control}

\author{\IEEEauthorblockA{\large{Sergio~Rozada, Dimitra~Apostolopoulou, and Eduardo~Alonso} \\ City, University of London \\ London, UK EC1V 0HB \\ Email: \{Sergio.Rozada, Dimitra.Apostolopoulou, E.Alonso\}@city.ac.uk}}

\maketitle
\begin{abstract}
The rise of microgrid-based architectures is heavily modifying the energy control landscape in distribution systems making distributed control mechanisms necessary to ensure reliable power system operations. In this paper, we propose the use of Reinforcement Learning techniques to implement load frequency control without requiring a central authority. To this end, we approximate the optimal solution of the primary, secondary, and tertiary control with the use of the Multi-Agent Deep Deterministic Policy Gradient (MADDPG) algorithm. Generation units are characterised as agents that learn how to maximise their long-term performance by acting and interacting with the environment to balance generation and load in a cost efficient way. Network effects are also modelled in our framework for the restoration of frequency to the nominal value. We validate our Reinforcement Learning methodology through numerical results and show that it can be used to implement the load frequency control in a distributed and cost efficient way.
\end{abstract}


\section{Introduction}

Electrical systems are undergoing major changes; there is a large number of deployed distributed generation that is slowly substituting large electromechanical generators~\cite{singh2014}. In the past, the majority of the load was met by large generation units, such as coal or nuclear plants. Nowadays, every single house can be a prosumer, i.e., produce and consume energy; and deliver excess energy to the network. This is facilitated by new market designs, e.g., peer-to-peer markets. 

This paradigm shift is shaping our understanding of energy and bringing us a whole new branch of opportunities as well as challenges. In this context of decentralisation, coordination amongst generators to balance generation and load~\cite{wang2012},~\cite{8485750} is more challenging. Traditionally, a hierarchical control system is used to meet this objective, i.e., primary, secondary and tertiary frequency control. Primary control keeps frequency between some acceptable limits; secondary control restores frequency to the nominal value; and tertiary control performs so in a cost efficient way. Secondary and tertiary control layers need a central authority to send appropriate control signals to generators to shift their generation to meet load. However, in this new paradigm where there are numerous generators participating in frequency control, the centralised approach has central limitations in terms of computation and privacy concerns. In this regard, new distributed schemes are necessary to deal with the aforementioned challenges~\cite{cady2017}. 

Different approaches have attempted to tackle this problem by implementing the traditional hierarchical control in a distributed manner (see, e.g.,~\cite{cady2017},~\cite{apostolopoulou2015},~\cite{apostolopoulou2015b}). In~\cite{guerrero2011}, the authors propose a methodology for primary control to mimic droop control strategies which are by nature decentralised algorithms that act upon each generator. The proposed methodology explicitly represents the modified system dynamics of having electronic inverters instead of large turbines. Moreover, efforts have been made to implement a decentralised secondary control scheme, e.g., the centralised averaging PI (CAPI) presented in~\cite{shafiee2014} and distributed averaging PI (DAPI) given in~\cite{simpson2015}. These algorithms use weighted averages of the frequency as the integral feedback. Despite their theoretical appeal, they suffer from lack of robustness, and their communication demands make them difficult to implement in real-life scenarios~\cite{dorfler2016}. {\color{black} Recently, several nature-inspired optimisation algorithms have been proposed to solve the primary and secondary layers of the load frequency control problem. Some of the most relevant ones are the water cycle algorithm (WCA) (see, e.g.,~\cite{latif2019comparative},~\cite{el2016water}), the yellow saddle goatfish (YSGA) (see, e.g.,~\cite{latif2020illustration}) and the butterfly optimisation algorithm (BOA) (see, e.g.,~\cite{latif2020optimum}). However, none of these techniques take the economic cost into consideration. 
Regarding tertiary control, it is usually common to solve a primal-dual algorithm that converges to the solution of the dual problem (see, e.g.,~\cite{7163587},~\cite{apostolopoulou2015},~\cite{yang2015minimum},~\cite{trip2018passivity},~\cite{moayedi2016}) where the communication between nodes enables joint global actions.} Nevertheless, as with other approaches, communication is intense between nodes and the system may become too complex. Multi-Agent Reinforcement Learning (MARL) is a promising alternative to implement load frequency control in a decentralised way addressing the aforementioned challenges (see, e.g.,~\cite{daneshfar2010load},~\cite{eftekharnejad2007stability}). The main drawback of these methods is their computational complexity, that grows exponentially in the number of agents. However, the rise of Deep Learning has opened the door to new techniques and algorithms that address these scalability issues in the load frequency control problem (see, e.g.,~\cite{rozada2020load},~\cite{yan2018data}). 

In MARL, various software agents learn optimal policies by negotiating, cooperating, and/or competing~\cite{sutton1988}. In this work we formulate the primary, secondary and tertiary control layers as an MARL problem so that the agents, i.e., generation units, learn to keep generation and load balanced in a cost efficient way by controlling the energy supply while minimising information exchange. More specifically, we recast the load frequency control problem as a Markov Decision Process (MDP), as is usually the practise in reinforcement learning problems. We define the states which are the frequency deviation, the control action of each generator, and their action space. We model the dynamic behaviour of the generators and the network to determine the probability state transition function of the MDP. We design the reward function of the agents so that frequency deviation and total cost are minimised. The design of the reward function is critical, since it determines the behaviour that each agent will learn. In order to determine the reward function we make use of the frequency deviation as well as optimality conditions of the economic dispatch problem to incorporate the cost component in the proposed framework. We use this setup to estimate the action-value function of each state-action pair with the Multi-Agent Deep Deterministic Policy Gradient (MADDPG) algorithm. MADDPG is an actor-critic algorithm; this means that the architecture of each agent is split into two. First, the actor directly estimates an action while secondly, the critic assesses the suitability of an action by estimating the action-value function of the state-action pair. In MADDPG, the critics use central information to teach each actor the dynamics of the environment as well as the behaviour of the rest of the agents. In operation, actors only use local information since they have learnt how other actors behave during the training phase. Each actor and critic is modelled with a Long Short-Term Memory Network (LSTM) so that previous history is stored and acted upon. {\color{black}To summarise, the contributions of the paper are: i) reformulation of the load frequency control problem as a Markov Decision Process (MDP); ii) use of a detailed model taking into account the network, renewable-based generation and generator rate constraints; iii) design of the reward function of the agents so that frequency deviation and total cost are minimised; iv) development of the proposed framework to solve the optimal load frequency control in a fully distributed manner with only the use of local information; and v) validation of its robustness against uncertainty introduced from renewable-based generation. This problem was initially introduced in~\cite{rozada2020load} and is extended in this paper to implement tertiary control or economic dispatch, i.e., the generation units modify their output to meet the change in load in a cost efficient way; to include a detailed power systems' model by explicitly incorporating the network, wind-based generation and a more realistic model of synchronous generator by including the generator rate constraints (GRC).}

The remainder of the paper is organised as follows. In Section~\ref{sec2} we describe the power system model that we adopt to develop our analysis framework. In Section~\ref{sec3}, we formalise the frequency control problem as an MARL problem. In Section~\ref{sec4}, MADDPG is used to implement primary, secondary and tertiary control in a multi-agent problem. In Section~\ref{sec5}, we present numerical studies to demonstrate that the proposed methodology is a valid alternative to solve load frequency control in a distributed and cost efficient manner. In Section~\ref{sec6}, we summarise the results and make some concluding remarks.

\section{Preliminaries}
\label{sec2}

In this section, we introduce the secondary and tertiary control models that we utilise to develop our framework. More specifically, we introduce dynamic models for synchronous generators, the automatic generation control (AGC) system, the network, and the economic dispatch.

The system frequency indicates if supply and demand are properly balanced. When the generated power exceeds the load the system frequency increases. Similarly, the system frequency decreases if generation is not sufficient to meet the load. Thus, controlling the frequency of the system is a standard approach to balance demand and supply~\cite{Wood:1996}. The frequency control is divided in a hierarchy of three layers: primary, secondary and tertiary control. In primary control, generation and demand are rapidly balanced since the synchronous generators are either speeding up or slowing down due to the load generation imbalance. This is achieved by a decentralised proportional control mechanism called droop control~\cite{book:sauer_pai_2007}. Then, a secondary control layer implements an integral control that compensates the steady-state error derived from droop control. Automatic Generation Control (AGC)~\cite{glavitsch1980} implements the secondary control layer collecting information from all generation units in a centralised way. Last, the tertiary control layer is related to the economic aspect of power system operations. This layer establishes the load sharing between the sources so that the operational costs are minimised~\cite{kirschen2004}. Tertiary control is implemented through the economic dispatch, which calculates the optimal operating point in an offline process. Next, we present two models, i.e., Model I and Model II, for the description of the power system dynamics. These two models will be used to formulate the frequency control problem of a power system with $n$ generators denoted by $\mathscr{G} = \{G_1,\ldots,G_n\}$.

\subsection{Model I: Balancing Authority (BA) Area Model}
\label{mod2}
It is common in power systems operations to model the dynamic behaviour of the entire balancing authority area instead of individual generators. In this regard, we define by $\Delta \omega$ the deviation of the centre of inertia speed from the synchronous speed; the total mechanical power produced $P_{SV} = \sum_{i \in \mathscr{G}}P_{SV_i}$, with $P_{SV_i}$ the mechanical power of generator $i$; and the total secondary command $Z_G = \sum_{i \in \mathscr{G}} z_i$, with $z_i$ the participation of generator $i$ to AGC. Then the BA area dynamics are:
\begin{eqnarray}
\label{eq:swing}
M\frac{d\Delta \omega}{dt} &=& P_{SV}-P_{G}-D \Delta \omega, \\
\label{eq:PG}
 T_{SV} \frac{d P_{SV}}{dt} &= &-P_{SV}+Z_G-\frac{1}{R_{D}} \Delta \omega, 
 \end{eqnarray} 
\noindent where $M = \frac{2 H}{\omega_s}$, with $H$ the system inertia constant and $\omega_s$ the synchronous speed; $T_{SV} = \frac{\sum_{i \in \mathscr{G}} T_{SV_i}}{n}$, $T_{SV_i}$ the time constant of the mechanical power dynamics of generator $i$; $D=\sum_{i \in \mathscr{G}}D_i$, with $D_i$ the machine $i$ damping coefficient; $\frac{1}{R_D} = \sum_{i \in \mathscr{G}} \frac{1}{R_{D_i}}$ with $R_{D_i}$ the governor droop of generator $i$. In this case we neglect the network effects and set $P_G = P_L(1+\rho)$, where $P_L$ is the system load and $\rho$ denotes the sensitivity of the losses with respect to the system load. The normalised participation factor of bus load changes $\Delta P_{L_i}$ with respect to total system load change $\Delta P_L$ is denoted by $\sigma_i$, the output of generator $i$ is denoted by $P_i$, then $\rho$, which denotes the sensitivity of the losses with respect to the system load is
\begin{equation}
\rho= \sum_{i \in \mathscr{G}} \sigma_i \frac{\partial P_\text{losses}}{\partial P_{i}}.
\end{equation}

\subsection{Model II: Synchronous Generator Dynamics }

\label{mod1}
In Model II the individual generators' dynamics are represented. For the $i^\text{th}$ synchronous generator, the three states are the rotor electrical angular position $\delta_i$, the deviation of the rotor electrical angular velocity from the synchronous speed $\Delta \omega_i$, and the mechanical power $P_{SV_i}$.  We denote by $z_i$ the participation of each generator $i$ in AGC. The evolution of the three states of the generator $i$ is determined by:
\begin{eqnarray}
 \label{eq:rotor_change}
 \frac{d\delta_i}{dt} &=& \Delta \omega_i, \\
 \label{eq:swing_i}
 M_i\frac{d\Delta \omega_i}{dt} &=& P_{SV_{i}}-P_{i}-D_i \Delta \omega_i, \\
 \label{eq:PG_i}
 T_{SV_i} \frac{d P_{SV_i}}{dt} &= &-P_{SV_i}+z_{i}-\frac{1}{R_{D_i}} \Delta \omega_i, 
 \end{eqnarray}
\noindent where the inertia constant is $H_i$; the synchronous speed is $\omega_s$ and $M_i = \frac{2 H_i}{\omega_s}$; the machine damping coefficient is $D_i$; the governor droop is $R_{D_i}$; and the parameter $z{_i}$ is an input provided by the AGC. The definitions of the machine parameters may be found in~\cite{book:sauer_pai_2007}. The output of generator $i$ $P_i$ is determined by \eqref{network_effects}.

\subsection{Network}

Let us consider a power system with $N$ nodes and $P_{L_i}$ represents the real power load at bus $i$.  Further, let $Q_{i}$ and $Q_{L_i}$ denote the reactive power supplied by the synchronous generator and demanded by the load at bus $i$, respectively. Then, we model the network using the standard nonlinear power flow formulation (see, e.g., \cite{book:sauer_pai_2007}); thus, for the $i^\text{th}$ bus, we have that
 \begin{eqnarray}
 \label{eq:active_power}
 P_i-P_{L_i} =V_i \sum_{k=1}^N V_k(G_{ik}\cos \theta_{ik} +B_{ik}\sin \theta_{ik}), \\
 \label{eq:reactive_power}
 Q_i -Q_{L_i}=V_i \sum_{k=1}^N V_k(G_{ik}\sin \theta_{ik} -B_{ik}\cos \theta_{ik}),
 \end{eqnarray}
 \noindent where $G_{ik}+jB_{ik}$ is the $(i,k)$ entry of the network admittance matrix and $\theta_{ik} = \theta_{i}-\theta_{k}$. 
 
We assume that i) bus voltage magnitudes are $|V_i|=1$p.u for $i = 1,\dots,N$, ii) lines are lossless and characterised by their susceptances $B_{ik}=B_{ki}>0$ for $i,k = 1,\dots,N$ with $i \neq k$, iii) reactive power flows do not affect bus voltage phase angles and frequencies, and iv) coherency between the internal and terminal voltage phase angles of each generator so that these angles tend to \textit{swing together}, i.e., $\delta_i = \theta_i$. As a result, we neglect \eqref{eq:reactive_power} and simplify \eqref{eq:active_power} to be:
 \begin{equation}
 \label{network_effects1}
 P_i - P_{L_i} = \sum_{\substack{k=1 \\ i\neq k}}^{N} B_{ik}(\delta_i - \delta_k).
 \end{equation}
 \noindent If bus $i$ does not contain a generator then $P_i=0$.
 
 In order to increase the accuracy of \eqref{network_effects1} we can slightly modify it by incorporating an approximation of the losses. We define the normalised participation factor of bus load changes $\Delta P_{L_i}$ with respect to total system load change $\Delta P_L$ by $\sigma_i$, then $\rho_i$, which denotes the sensitivity of the losses with respect to the system load at bus $i$ is
\begin{equation}
\rho_i =  \sigma_i \frac{\partial P_\text{losses}}{\partial P_{i}}.
\end{equation}
\noindent Then \eqref{network_effects1} becomes:
 \begin{equation}
 \label{network_effects}
 P_i - (1+\rho_i)P_{L_i} = \sum_{\substack{k=1 \\ i\neq k}}^{N} B_{ik}(\delta_i - \delta_k).
 \end{equation}

 \subsection{Economic Dispatch}

The economic dispatch process is formulated as an optimisation problem, where the objective function that needs to be minimised is the sum of the individual costs of all generating units, $c_i(P_i)$, for $i \in \mathscr{G}$; this is typically a quadratic function that computes the production cost of each generation unit. Here, the constraint is that the system has to keep generation and load balanced; if generation and load are balanced then frequency is also nominal. The economic dispatch problem may be formulated as:
\begin{equation}
 \begin{aligned}
       &\underset{P_{i}}{\text{minimize }} \sum_{i \in \mathscr{G}} c_i(P_{i})  \\
       &\mbox{subject to } \sum_{i \in \mathscr{G}} P_{i} = (1+\rho)P_L.
 \end{aligned}
  \label{eq:ed}
  \end{equation}
\noindent 

{
\color{black}

\subsection{Wind-based generation}

The increasing penetration of renewable-based resources in the system introduces a source of uncertainty in power system operations and thus in load frequency control problems. In this regard, we investigate the effect of wind-based generation units in the proposed framework. The relationship between the wind speed and the generated power can be efficiently modelled as a linear dynamical system~\cite{apostolopoulou2013effects}. More specifically, $P_W$ denotes the real wind generation power output; $\Delta v$ the variation of the wind speed; $\alpha_{W_1}$ and $\alpha_{W_2}$ are parameters that depend on the wind turbine characteristics; $W_t$ is a Wiener process and $\beta_{W_1}$ and $\beta_{W_1}$ coefficients that represent prior knowledge of the wind speed probability distribution. Then, the dynamics of the wind-based generation power output are formulated as follows:
 \begin{eqnarray}
 \label{eq:wind_power}
 \frac{d \Delta P_W}{dt} = \alpha_{W_1} \Delta P_W + \alpha_{W_2} \Delta v, \\
 \label{eq:wind_speed}
 d \Delta v = \beta_{W_1} \Delta v dt + \beta_{W_2} d W_t.
 \end{eqnarray}
}

 \section{Multi-Agent Reinforcement Learning for Load Frequency Control}
 \label{sec3}
 
\begin{figure}[t!]
\centering
\includegraphics[width=0.44\textwidth]{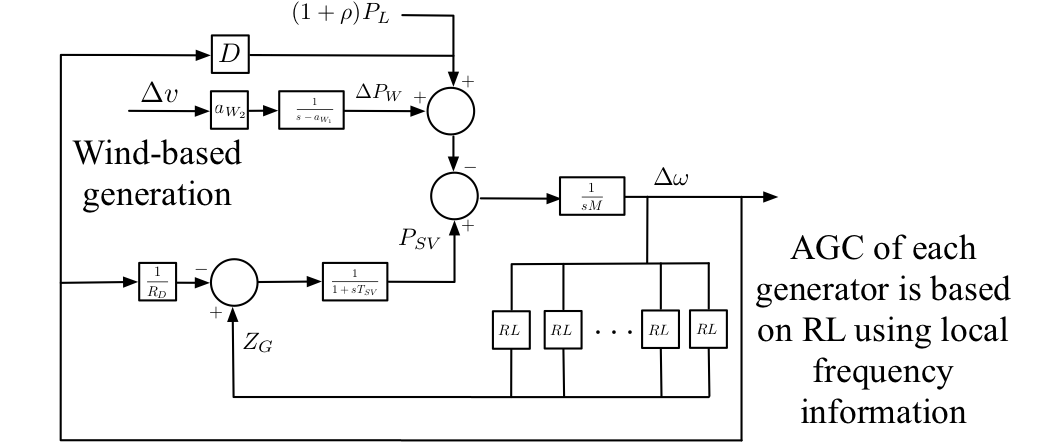} 
\caption{Diagram of proposed load frequency control scheme.}
\label{fig_block1}
\end{figure}

In this section we  formulate the load frequency control problem as an MARL problem. Reinforcement Learning (RL) is an area of Machine Learning strongly related with the notion of software agents~\cite{nwana1996}. RL studies how software agents interact in an environment to maximize their long-term performance. We use MARL to train a collection of agents how to implement the load frequency control problem in a distributed way. {\color{black} In this paper, an agent physically represents the controller of a generation unit. As such, by using the MARL scheme, which allows for a fully distributed control architecture, the load frequency control can be achieved with no communication infrastructure between the agents, i.e., generating units. The controller physical circuit of each generator does not have to be connected with any of the other controllers, thus allowing a physical ``distribution'' of the load frequency control system. A diagram of the entire architecture and how the reinforcement learning based AGC design fits in the power system dynamics is shown in Fig.~\ref{fig_block1}.}

RL problems are mathematically formalized through a Markov Decision Process (MDP)~\cite{littman1994}, that is defined as the tuple:
 \begin{equation}
 MDP = \langle S,A,P,R \rangle,
 \end{equation}
\noindent where each term is:
\begin{itemize}
\item \textit{$S$ or state space}: all possible states where the agent can be in the environment. There are two continuous states in the load frequency control: the deviation from synchronous speed which is quantified by $\Delta \omega_i$ for each generator $i$ or $\Delta \omega$ in the case of the BA area model; and $z_i$, the current control action of each generator $i$. These states provide to the agent information about the difference between demand and supply and how much they are contributing to the total generation.
\item \textit{$A$ or action space}: all possible actions that each agent takes in every state. Our agents-generators can increase or decrease the control action $z_i$ in order to modify the state of the environment.
\item \textit{$P$ or probability state transition function}: it defines the dynamics of the environment, modelling the transition between states. This is determined by the modelling approach we are following, i.e., Model I as described in Section~\ref{mod2} or Model II given in Section~\ref{mod1}. For the BA area model or Model I described in Section~\ref{mod2} the transition equations derived from \eqref{eq:swing} and \eqref{eq:PG} are:
\begin{eqnarray}
&M\frac{d\Delta \omega^\text{new}}{dt} = P_{SV}^\text{old}-(1+\rho)P_{L}-D \Delta \omega^\text{old}, \\
 &T_{SV} \frac{d P_{SV}^\text{new}}{dt} = -P_{SV}^\text{old}+Z_{G}^\text{new}-\frac{1}{R_{D}} \Delta \omega^\text{old}, \\
 &Z_{G}^\text{new} = \sum_{i \in \mathscr{G}} z_{i}^\text{new}, \\
 &z_{i}^\text{new} = z_{i}^\text{old}+\Delta z_i, \\
&\Delta \omega^\text{new} = \Delta \omega^\text{old}+ \frac{d\Delta \omega^\text{new}}{dt} \Delta t,\\
&P_{SV}^\text{new}= P_{SV}^\text{old}+ \frac{dP_{SV}^\text{new}}{dt} \Delta t.
 \end{eqnarray}

For the detailed modelling of Model II given in Section~\ref{mod1} the transition equations based on the  \eqref{eq:rotor_change},\eqref{eq:swing_i}, \eqref{eq:PG_i} and \eqref{network_effects} are as follows:
\begin{eqnarray}
&\frac{d\delta_i^\text{new}}{dt} = \Delta \omega_i^\text{old}, \\
&M_i\frac{d\Delta \omega_i^\text{new}}{dt} = P_{SV_i}^\text{old}-P_i-D_i \Delta \omega_i^\text{old}, \\
&T_{SV_i} \frac{d P_{SV_i}^\text{new}}{dt} = -P_{SV_i}^\text{old}+z_{i}^\text{new}-\frac{1}{R_{D_i}} \Delta \omega_i^\text{old}, \\
&z_i^\text{new} = z_i^\text{old}+\Delta z_i, \\
&\Delta \omega_i^\text{new} = \Delta \omega_i^\text{old}+ \frac{d\Delta \omega_i^\text{new}}{dt} \Delta t,\\
&\delta_i^\text{new} = \delta_i^\text{old} + \frac{d\delta_i^\text{new}}{dt} \Delta t, \\
&P_{SV_i}^\text{new} = P_{SV_i}^\text{old}+ \frac{dP_{SV_i}^\text{new}}{dt} \Delta t,\\
&P_i - (1+\rho_i)P_{L_i} = \sum_{\substack{k=1 \\ i\neq k}}^{N} B_{ik}(\delta_i^\text{new} - \delta_k^\text{new}),
\end{eqnarray}
\noindent where $\Delta z_i$ is the increase or decrease in power generation by each unit $i$ in $\mathscr{G}$ estimated by each agent. MADDPG is used to estimate $\Delta z_i$, as described in Section \ref{sec4}.

\item \textit{$R$ or reward function}: it defines a numerical signal or reward expressing the value of being in a state and performing an action. The reward function considers two different dimensions in our case: frequency deviation and operational costs.
\end{itemize}

MARL attempts to learn an optimal policy $\pi:S\mapsto A$ that maximises the cumulative reward, or return. However, the reward is instantaneous and does not address the global nature of the task, i.e., one bad action can lead to an extremely good position from which the agent can obtain a good reward. Thus, action-value functions $Q^\pi$ are used in RL to express the expected long-term reward achievable from being in an state, taking an action and following a policy $\pi$:
\begin{equation}
Q^\pi(s_t,a_t) = \mathbb{E}_\pi \left[R_t|s_t,a_t\right] =\mathbb{E}_\pi \left[\sum_{k=0}^\infty \gamma^k r_{t+k+1}|s_t,a_t\right] , 
\end{equation}
\noindent where $\mathbb{E}[\cdot]$ is the expectation operator, $\gamma$ is the discount factor, which expresses the fidelity in long-term predictions of $Q^\pi$, the cumulative reward achievable in the long run $R_t$, and the reward $r_{t}$ at time $t$. Most algorithms strongly support their learning process in value functions, such as Q-learning~\cite{watkins1992}.

The action-value function associates a value $Q^\pi$ to each state-action pair. However, when the number of states and actions is too large, it becomes computationally challenging to estimate them efficiently. Recent work has merged the field of RL with Deep Learning, giving birth to a powerful algorithm called Deep Q-learning (DQN)~\cite{mnih2013}. This algorithm uses deep neural networks as parametric function approximators to estimate the action-value function of each state-action pair.

The spectrum of existing algorithms to solve MARL problems is wide. Most of them use game-theoretic approaches to augment Q-learning, i.e., Nash Q-learning or minimax Q-learning~\cite{bosoniu2008}. In our problem, state and action spaces are continuous and the interaction of various agents is required. This limits the range of algorithms available in the literature. MADDPG addresses both problems at the same time~\cite{lowe2017}.

\section{Multi-Agent Deep Deterministic Policy Gradient}
\label{sec4}

In this section, we present the selection of the appropriate multi-agent actor-critic algorithm that takes into account the fact that state and action spaces are continuous, and the design of the reward function. 

MADDPG is an actor-critic algorithm. This means that the architecture of each agent, or generation unit, is split into two. First, the actor directly estimates an action while secondly, the critic assesses the value of the action by estimating the action-value function $Q^\pi$ of the state-action pair. The $Q^\pi$ estimated by the critic is used by both the critic and the actor to learn how to behave in the environment. In MADDPG, the critics use central information to teach each actor the dynamics of the environment as well as the behaviour of the rest of the agents. In operation, actors only use local information because they have learnt how other actors will behave.

\begin{figure}[t!]
\centering
\includegraphics[width=0.2\textwidth]{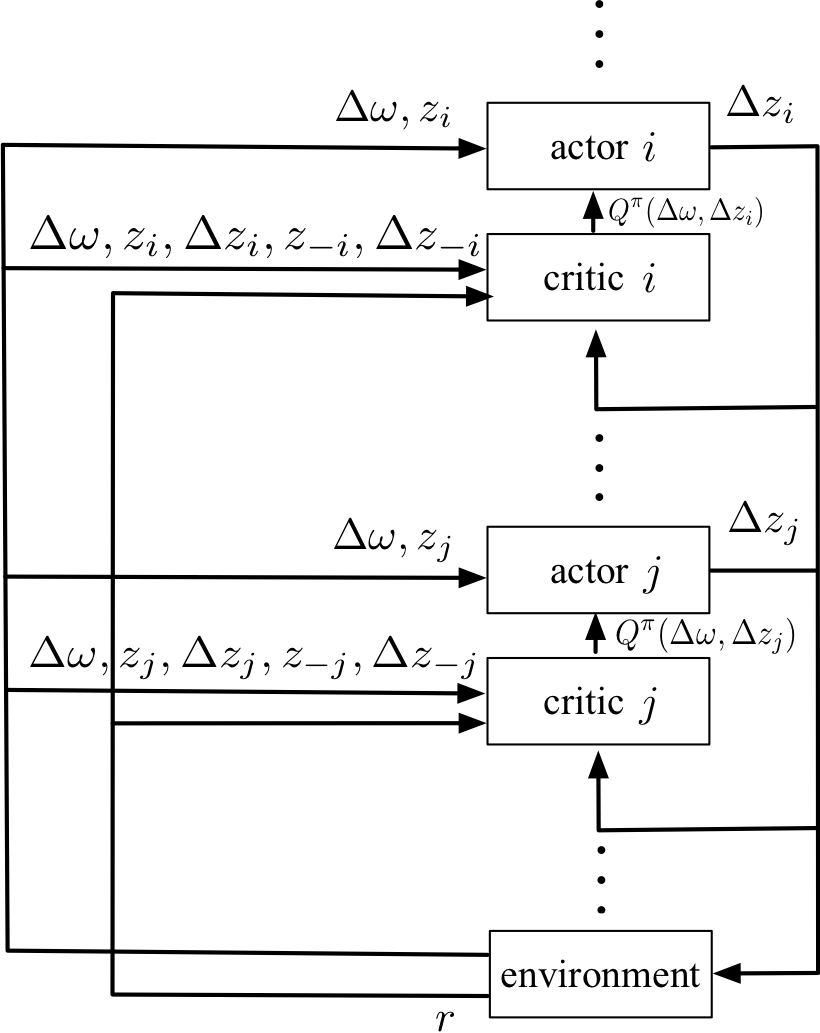} 
\caption{ MADDPG schema in a frequency control scenario.}
\label{fig1}
\end{figure}

We describe the actor-critic algorithm for the BA model or Model I. This is the same for the detailed modelling of Model II presented in Section~\ref{mod1}, the only difference is that instead of the deviation from the centre of inertia the input to the actor and the critic is the deviation of the rotor speed from the synchronous speed of each generator $\Delta \omega_i$. For the BA area model given in Section~\ref{mod2} we have each actor $i$ that estimates $\Delta z_i$ given the state of the environment $\Delta \omega$ and its current $z_i$. Each critic assesses each state-action pair defined by the environment and the actions of all the actors. The critic estimates each state-action action-value that is used during actor's training, as it can be seen in Fig.~\ref{fig1}. We denote by $\Delta z_{-i}$ ($\Delta z_{-j}$) the action predicted by all other actors besides $i$ ($j$) and $z_{-i}$ ($z_{-j}$) the control action state of all other actors besides $i$ ($j$).

 Deep Recurrent Neural Networks, particularly Long Short-Term Memory Networks (LSTMs) \cite{hochreiter1997}, are used to model each actor and critic. LSTMs implement memory so that previous history is stored and acted upon~\cite{lample2017}. In MDPs the Markov assumption states that the current state comprises all information needed to choose an action. However, in the frequency control problem the dynamics are quite complex and the Markov assumption may not hold. Thus, LSTMs help us correcting the violation of the Markov assumption.
 
 \begin{figure}[b!]
\centering
\includegraphics[width=0.4\textwidth]{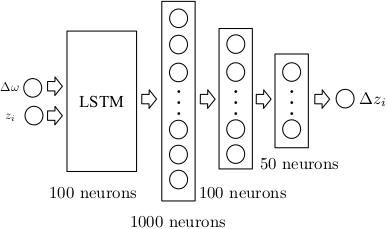} 
\caption{Architecture of the MADDPG actor.}
\label{fig2}
\vspace{-1\baselineskip}
\end{figure}

 \begin{figure}[t!]
 \vspace{-1\baselineskip}
\centering
\includegraphics[width=0.4\textwidth]{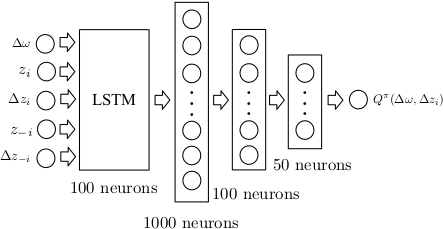} 
\caption{Architecture of the MADDPG critic.}
\label{fig3}
\end{figure}

The actor network, see Fig.~\ref{fig2}, has as inputs $\Delta \omega$ and $z_i$ and computes $\Delta z_i$. The critic network, depicted in Fig.~\ref{fig3}, has as inputs the frequency state of the network $\Delta \omega$; the secondary control action $z_i$; the change in the action predicted by the actor associated to that critic $\Delta z_i$; the secondary control action $z_{-i}$; and the change in the action predicted by all other actors $\Delta z_{-i}$. The critic network then computes the $Q^\pi(\cdot)$ value of the state-action pair estimated by the actor associated with that critic. As seen in the respective figures, both networks have a $100$-neuron LSTM that implements memory, and three more $1000$, $100$ and $50$ fully-connected hidden layers. {\color{black} Generation rate constraints can be easily introduced in this neural network based approach. More specifically, the output $\Delta z_i$ of each actor can be bounded by applying a non-linear function (e.g., sigmoid function, hyperbolic tangent, etc.) at the output of the network, so the agent has to learn that it cannot generate at an unrealistic rate.}

The design of the reward function is critical, since it determines the behaviour that agents will learn. Ideally, the reward function incorporates two different components: (i) the frequency state of the environment to solve the primary and secondary problem; and (ii) the operational cost associated with the system to solve the tertiary control problem. Taking into account the frequency component in the reward function is straightforward since we set a higher reward for smaller frequency deviations. Next, we need to determine how to define the reward function in order to take into account the cost component. In this regard, we study the case where the cost functions of generators are of the form $c_i(P_i) = a_i P_i^2+\beta_i P_i+\gamma_i$ for $i \in \mathscr{G}$~\cite{apostolopoulou2015}. The cost minimisation is part of the tertiary control in the hierarchical control setting; the formulation of which may be found in \eqref{eq:ed}. For quadratic cost functions under no generation limits we may find the optimal solution in an analytical way~\cite{Wood:1996}. The Lagrangian may be written as
\begin{equation*}
\mathcal{L}(P_{i},\lambda)=\sum_{i \in \mathscr{G}} c_i(P_{i})+\lambda \left( (1+\rho)P_L-\sum_{i \in \mathscr{G}} P_{i}\right),
\end{equation*}
\noindent where $\lambda$ is the dual variable of the power balance constraint. The necessary conditions for a minimum are
\begin{equation}
\frac{\partial \mathcal{L}}{\partial P_{i}}=0 \Rightarrow \frac{d c_i}{dP_i}-\lambda=0 \Rightarrow 2a_iP_i+\beta_i = \lambda, \forall i \in \mathscr{G}.
\label{eql}
\end{equation}
The solution to the problem above defines the base point operation of tertiary control. We now define with the aid of participation factors how would a generator participate in a load change so that the new load is served in a cost efficient way. 
We start from a given base point $\lambda_0$ as found from \eqref{eql}. Assume the change in load is $\Delta P_L$; the system incremental cost moves from $\lambda^0$ to $\lambda^0+\Delta \lambda$. For a small change in power output on unit $i$ $\Delta P_i$ we have
\begin{equation}
\Delta \lambda \approx \frac{d^2c_i}{dP_{i}^2} \Delta P_{i} \Rightarrow  \Delta P_{i} = \frac{\Delta \lambda}{\frac{d^2c_i}{dP_{i}^2}}, \forall i \in \mathscr{G}.
\end{equation}
\noindent Thus we wish that each generator $i$ changes its output so the following holds
\begin{equation}
\Delta \lambda = \frac{d^2c_i}{dP_{i}^2}  \Delta P_{i} = \frac{d^2c_{j}}{dP_{j}^2}  \Delta P_{j}, \forall i,j \in \mathscr{G},
\end{equation}
\noindent i.e., for each generator the change in the action $\Delta z_i$, $i \in \mathscr{G}$ we wish that 
\begin{equation}
\left|\Delta z_i \frac{d^2c_i}{dP_{i}^2} -\Delta z_j \frac{d^2c_{j}}{dP_{j}^2}\right| =0, \forall i,j \in \mathscr{G}.
\end{equation}

Now we use these two conditions, i.e., frequency deviation and cost information, to determine the reward functions for each modelling approach.

\subsection{Reward function -- Model I}

We construct two conditions that will be used in the formulation of the reward function. The first condition is:
\begin{equation*}
C1: |\Delta \omega| < \epsilon_1,
\end{equation*}
\noindent where $\epsilon_1$ is some selected tolerance; this condition ensures that the reward function $r$ will reward actions that help in frequency restoration. The second condition is:
\begin{equation*}
C2: \frac{\sum_{i \in \mathscr{G}} \sum_{j \in \mathscr{G},j > i}\left|z_i \frac{d^2c_i}{dP_{i}^2} - z_j \frac{d^2c_{j}}{dP_{j}^2}\right|}{(n-1)!}< \epsilon_2,
\end{equation*}
\noindent where $\epsilon_2$ is some selected tolerance; this condition ensures that $r$ will reward actions that follow the cost efficient path. 

\begin{figure}[b!]
\vspace{-\baselineskip}
\centering
\includegraphics[width=0.48\textwidth]{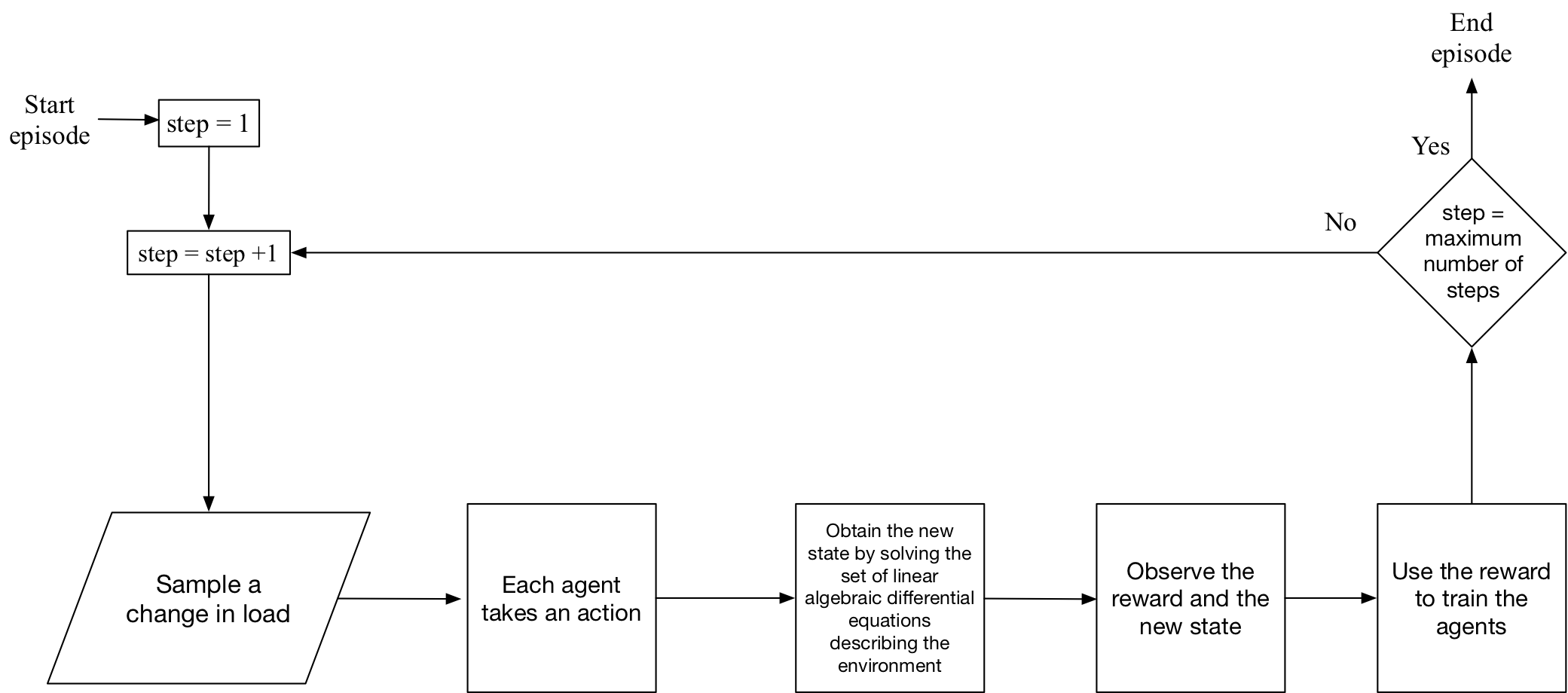} 
\caption{Diagram of the training process for developing the proposed load frequency control scheme.}
\vspace{\baselineskip}
\label{fig_block}
\end{figure}

When only the primary and secondary control problems need to be solved, the reward function may be formulated using $C1$ as
\begin{equation}
r = \begin{cases}
	d,& \text{if }C1,\\
	0, & \text{otherwise},
\end{cases},
\label{eqr_primary}
\end{equation}
\noindent where $d$ is a constant. 

On the other hand, by taking these two conditions into account we may formulate a general form of the reward function to solve all levels of control from primary to tertiary as
\begin{equation}
r = \begin{cases}
   d_1,& \text{if } C1 \land C2, \\
    d_2,              & \text{if } C1  \lor C2 ,\\
    0, & \text{otherwise},
\end{cases},
\label{eqr}
\end{equation}
\noindent where $\land$ is the logical \textit{and}; $\lor$ is the logical \textit{or}; $d_1$, $d_2$ are constants with $d_2<d_1$. This reward function both helps in frequency restoration as well as performs the latter in a cost efficient way, since the critic values actions higher that if both purposes are met.

\subsection{Reward function -- Model II}

In order to ensure frequency restoration, i.e., secondary control, we wish that $|\Delta \omega_i|< \epsilon$ for all $i \in \mathscr{G}$. To this end, we formulate the reward function as follows:
\begin{equation}
r= \begin{cases}
	d'_1,&  \exists i: |\Delta \omega_i|< \epsilon\\
	d'_2,&  \exists i,j: j \neq i,|\Delta \omega_i | \land |\Delta \omega_j |< \epsilon,  \\
	d'_3,&  \exists i,j,j': j \neq  j' \neq i, |\Delta \omega_i| \land |\Delta \omega_j|  \land | \Delta \omega_{j'}|< \epsilon,\\
	\vdots & \\
	d'_n,&  |\Delta \omega_1| \land |\Delta \omega_2| \land \dots \land | \Delta \omega_{n}|< \epsilon,\\
	0, & \text{otherwise}
\end{cases},
\label{eqr_primary2}
\end{equation}
\noindent where $\land$ is the logical \textit{and} sign; $d'_1,d'_2,\dots ,d'_n$ are constants with $d'_1<d'_2<d'_3<\cdots<d'_n$. This formulation ensures that the reward is higher when more generators' frequency deviation is smaller than a specified tolerance. In this work we have not performed the tertiary control for Model I, which is part of future work. 

{\color{black}A flow diagram of the training process of determining the RL-based control for each agent is shown in Fig.~\ref{fig_block}.}

\section{Numerical Results}
\label{sec5}

\begin{figure*}[b!]
\vspace{-1\baselineskip}
        \centering
        \begin{subfigure}[b]{0.42\textwidth}
                \centering
                \includegraphics[width=\textwidth]{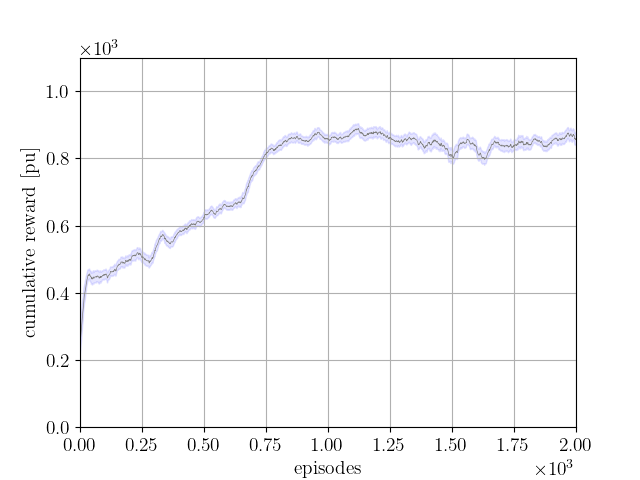}
                \caption{Smoothed cumulative reward per episode with 95\% confidence levels.}
                \label{figtrain}
        \end{subfigure} 
        \hspace{2pt}
          \begin{subfigure}[b]{0.42\textwidth}
                \centering
                \includegraphics[width=\textwidth]{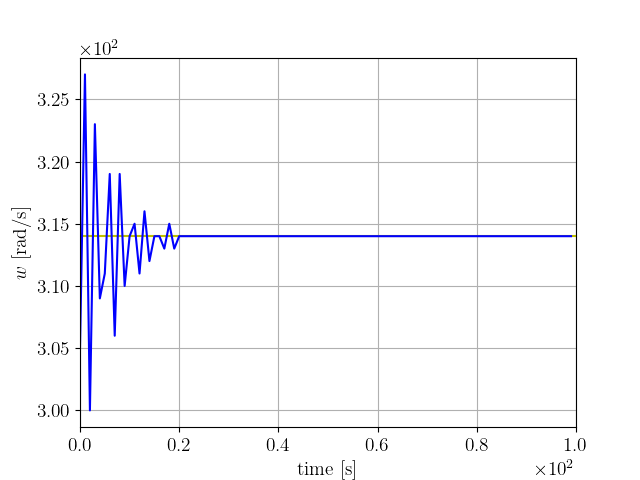}
                \caption{Centre of inertia speed.}
                 \label{fig4}
        \end{subfigure} \\
        \begin{subfigure}[b]{0.42\textwidth}
                \centering
                \includegraphics[width=\textwidth]{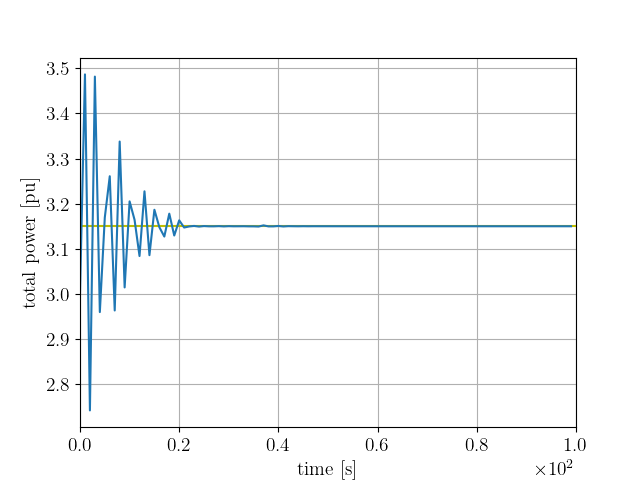}
                \caption{Power response of the eight agents.}
                \label{fig5}
        \end{subfigure} 
        \hspace{2pt}
          \begin{subfigure}[b]{0.42\textwidth}
                \centering
                \includegraphics[width=\textwidth]{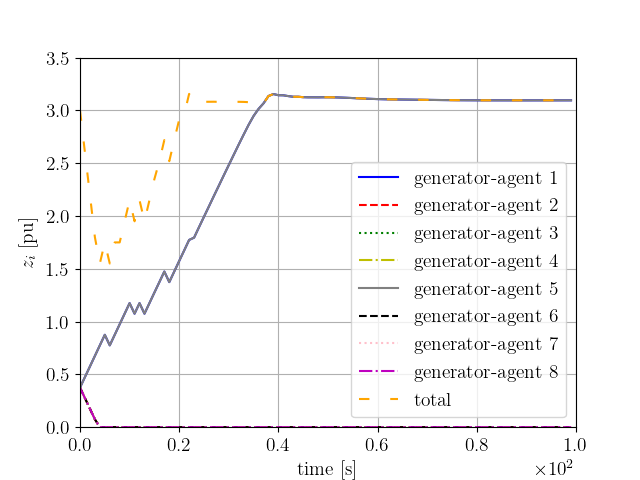}
                \caption{Secondary control action of the eight agents }
                 \label{fig6}
        \end{subfigure} 
        \caption{Secondary Control Model I: Change in load by $0.15$pu.}
          \label{figsub1}        
\end{figure*}

We validate the MARL methodology using three test systems. We formulate the reward function and present the results of the primary and secondary control problem for Model I and the detailed modelling of Model II taking into account the network effects. {\color{black} We also demonstrate the flexibility of the proposed methodology to incorporate generation rate constraints as well as its robustness against the uncertainty introduced due to wind-based generation}. Next, we formulate the reward function and present the results of all levels of control for one single BA area using Model I. We demonstrate that the generators are able to restore the system frequency back to nominal and operate at a point close to optimal when a change in load occurs in a distributed way. {\color{black} We compare the results with an standard distributed optimal load frequency control alternative ~\cite{7163587}.}

\subsection{Secondary Control -- Model I}

\begin {table}[t!]
\begin{center}
\begin{tabular}{cc}
 \hline
Nominal frequency & $f^\text{nom}=50$ Hz \\
\hline
Initial operating point & $P_i = 0.375$ pu, $i = 1,\dots, 8$  \\
\hline
Inertia parameter & $M = 0.1$ pu \\
\hline
Droop & $R_D = 0.1$ pu \\
\hline
Load damping & $D = 0.0160$ pu\\
\hline
Generator time constant & $T_{SV} = 30$ s\\
\hline
\end{tabular}
\end{center}
\caption{Eight-generator one-load power system data.}
\label{tabs1}
\end{table}

\begin{figure}[t!]
\centering
\includegraphics[width=0.18\textwidth]{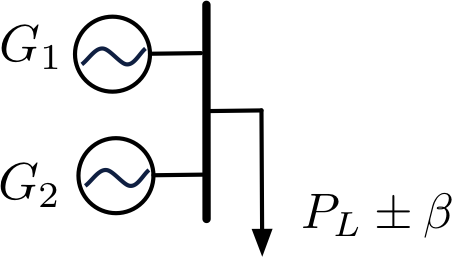} 
\caption{One-line diagram of a two-generator and one-load power system.}
\label{fig7}
\end{figure}

\begin {table}[t!]
\begin{center}
\begin{tabular}{cc}
 \hline
Nominal frequency & $f^\text{nom}=50$ Hz \\
\hline
Initial operating point & $P_1 = 1.5$ pu, $P_2 = 1.5$ pu \\
\hline
Inertia parameter & $M = 0.1$ pu \\
\hline
Droop & $R_D = 0.1$ pu \\
\hline
Load damping & $D = 0.0160$ pu\\
\hline
Generator time constant & $T_{SV} = 30$ s\\
\hline
\end{tabular}
\end{center}
\caption{Two-generator and one-load power system data.}
\label{tab1}
\end{table}

\begin{figure*}[t!]
        \centering
        \begin{subfigure}[b]{0.42\textwidth}
                \centering
                \includegraphics[width=\textwidth]{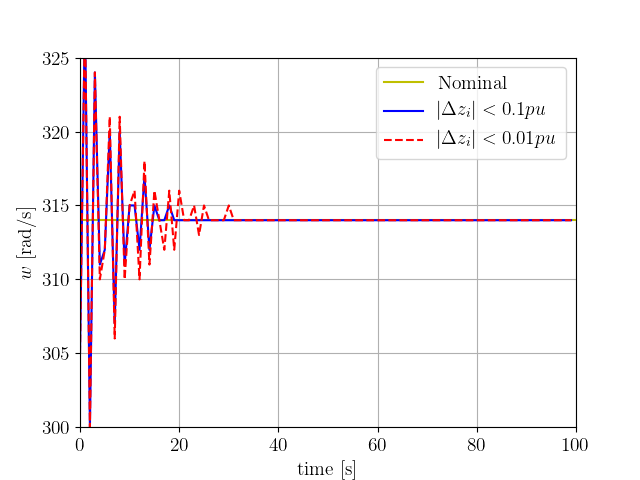}
                \caption{Centre of inertia speed.}
                \label{fig_fs_const}
        \end{subfigure} 
        \hspace{2pt}
          \begin{subfigure}[b]{0.42\textwidth}
                \centering
                \includegraphics[width=\textwidth]{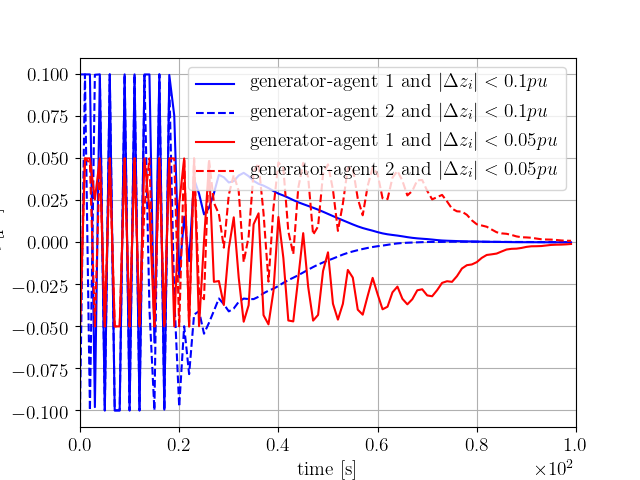}
                \caption{Secondary control action of the two agents.}
                 \label{fig_dz_const}
        \end{subfigure}    
        \caption{Secondary Control Model I: Change in load by $0.15$pu with $\Delta z_i$ bounded at different levels.}
          \vspace{-1\baselineskip}  
\end{figure*}

The test case to validate the secondary control using Model I comprises of a group of eight generating units or agents that interact with a load; the parameters of the environment can be found in Table~\ref{tabs1}. In each training episode, the load varies around a nominal set point randomly. The modification is indicated by $P_L\pm\Delta P_L = 3 \pm \beta$ pu, where $\beta$ follows a uniform distribution. The reward function has been derived following~\eqref{eqr_primary} and is defined by:
\begin{equation*}
r = \begin{cases}
   10,& \text{if } C1 \\
        0, & \text{otherwise}
\end{cases}.
\end{equation*}
During operation, only the actors interact with the environment. They only observe local information about the frequency of the system and the control action that they are executing. As a consequence of the training phase, they know how to act according to the states of the environment in order to keep load and generation balanced. The validation of the training is tested by changing the load by $0.15$ pu and observing how the generators modify their output.

In Fig.~\ref{figtrain} the cumulative reward obtained by the agents is depicted. The agents can obtain $1,000$ at maximum per episode, i.e., the maximum reward per step is $10$ and the number of steps per episode is $100$. The agents learn how to obtain higher rewards as the number of episodes increases, since if that was not the case the cumulative reward function would oscillate around small values near zero.

In Figs.~ \ref{fig4},~ \ref{fig5} the centre of inertia speed and power response of an eight-generator system when a single load increases by $0.15$ pu is depicted. However, as it can be seen in Fig.~ \ref{fig6}, the solution may be unrealistic given that the operational cost component is neglected in this test case. As such, one agent learns to balance the entire system while the others have zero output. Further analysis on secondary control using Model I may be found in~\cite{rozada2020load}.

\begin{figure*}[t!]
          \vspace{-1\baselineskip}  
        \centering
        \begin{subfigure}[b]{0.42\textwidth}
                \centering
                \includegraphics[width=\textwidth]{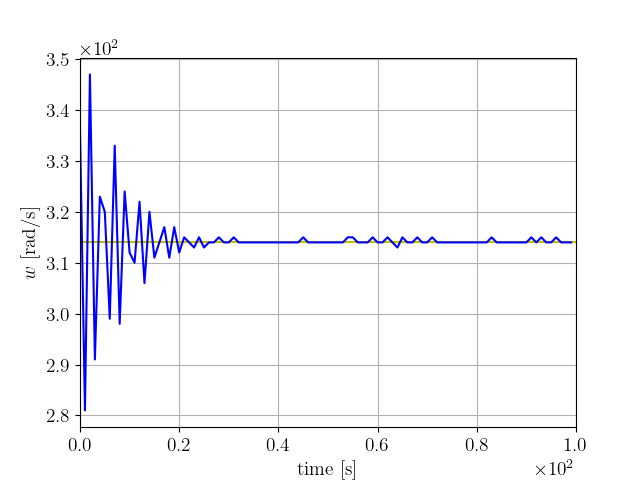}
                \caption{Centre of inertia speed.}
                \label{fig_fs_wien}
        \end{subfigure} 
        \hspace{2pt}
          \begin{subfigure}[b]{0.42\textwidth}
                \centering
                \includegraphics[width=\textwidth]{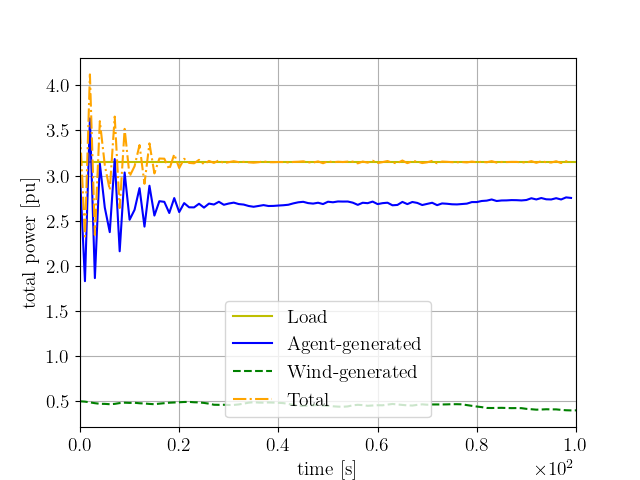}
                \caption{Secondary control action of the two agents.}
                 \label{fig_p_wien}
        \end{subfigure}    
        \caption{Secondary Control Model I: Change in load by $0.15$pu with wind-based generation.}
\end{figure*}

{ \color{black}

In order to highlight the ability of the proposed framework to implement generation rate constraints, we have used Model I to conduct numerical analyses of the response of a two-agent system, whose data can be found in Table~\ref{tab1} and is depicted in Fig.~\ref{fig7}, when the generation rate of each unit is bounded by different values. We modify the load by 0.15 pu and limit the output of each actor $\Delta z_i$ to $0.1$, $0.05$, and $0.01$ pu respectively by using a hyperbolic tangent function. As depicted in Fig.~\ref{fig_fs_const}, although the system manages to meet the new load, the generation rate constraints affect the elapsed time until the new steady state is reached. This can be also inferred by observing Fig. ~\ref{fig_dz_const}, where the actual values of $\Delta z_i$ are shown. When the generation rate is more constrained, i.e., the generators are allowed to modify their output in smaller increments, the system spends more time balancing generation and demand, as expected.

\begin{figure}[t!]
	\centering
	\includegraphics[width=0.3\textwidth]{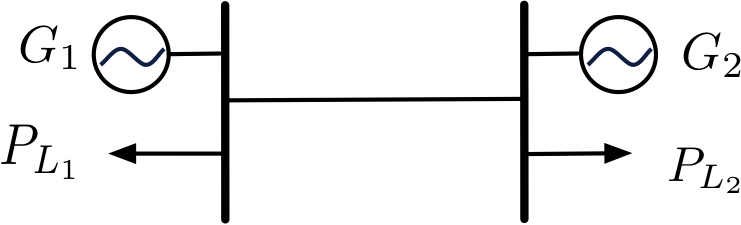} 
	\caption{One-line diagram of a two-generator and two-load power system.}
	\label{fig13}
	\vspace{-1\baselineskip}
\end{figure}

\begin {table}[t!]
\begin{center}
	\begin{tabular}{cc}
		\hline
		Nominal frequency & $f^\text{nom}=50$ Hz \\
		\hline
		Initial operating point & $P_1 = 1.5$ pu, $P_2 = 1.5$ pu \\
		\hline
		Inertia parameter & $M_1 = 0.1$ pu, $M_2=0.15$ pu\\
		\hline
		Droop & $R_{D_1} = 0.1$ pu, $R_{D_2} = 0.08$ pu \\
		\hline
		Load damping & $D_1 = 0.0160$ pu, $D_2 = 0.0180$ pu\\
		\hline
		Generator time constant & $T_{SV_1}, T_{SV_2} = 30$ s\\
		\hline
	\end{tabular}
\end{center}
\caption{Two-generator and two-load power system data.}
\label{tab2}
\end{table}

Next, we validate that the proposed framework is robust against the uncertainty introduced by wind-based generation, as described in Section~\ref{sec2}. We model a wind-based generator as a stochastic process with parameters $\alpha_{W_1}=-0.002$, $\alpha_{W_2}=0.01$, $\beta_{W_1}=-0.5$, and $\beta_{W_2}=-0.4$,. We train the two-agent system of Table~\ref{tab2} to balance the load under such conditions. In Figs.~\ref{fig_fs_wien},~\ref{fig_p_wien}, it can be seen that the load is met and the frequency is close to nominal under the scenario that the wind-based generation evolves randomly. More specifically, minor variations appear in the frequency response as the agents adapt to re-balance the load.

}

\subsection{Secondary Control -- Model II}

Analogously, we have designed a test case to validate the performance of the proposed solution using the detailed Model II. The dynamic behaviour of two generators that are part of a BA area is explicitly taken into account as well as the network. The configuration of the system that has two loads, i.e., $P_{L_1}$ and $P_{L_2}$, can be found in Fig.~\ref{fig13}. The parameters of the environment can be found in Table~\ref{tab2}. In each training episode, each load varies around a nominal set point randomly. The modification of each load is indicated by $P_{L_i}\pm\Delta P_{L_i} = 1.5 \pm \beta$ pu, where $\beta$ follows a uniform distribution.

\begin{figure*}[t!]
\vspace{-1\baselineskip}
        \centering
        \begin{subfigure}[b]{0.42\textwidth}
                \centering
                \includegraphics[width=\textwidth]{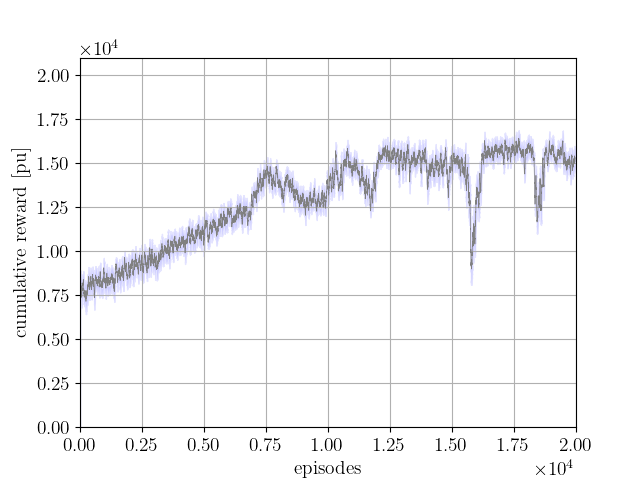}
                \caption{Smoothed cumulative reward per episode with 95\% confidence levels.}
                \label{fig14}
        \end{subfigure} 
        \hspace{2pt}
          \begin{subfigure}[b]{0.42\textwidth}
                \centering
                \includegraphics[width=\textwidth]{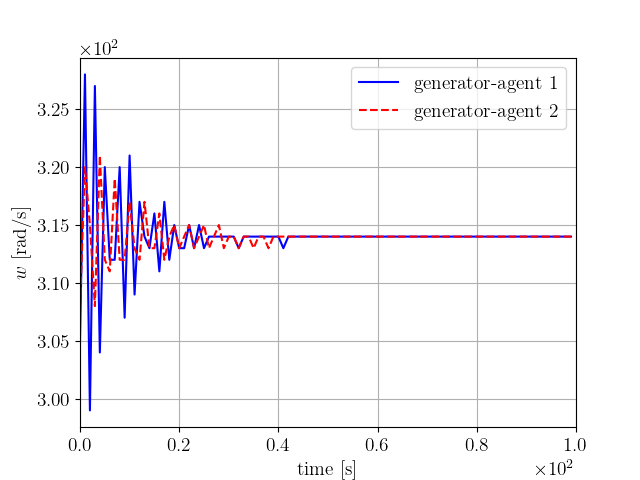}
                \caption{Rotor electrical angular velocity of the two generators.}
                 \label{fig15}
        \end{subfigure} \\
        \begin{subfigure}[b]{0.42\textwidth}
                \centering
                \includegraphics[width=\textwidth]{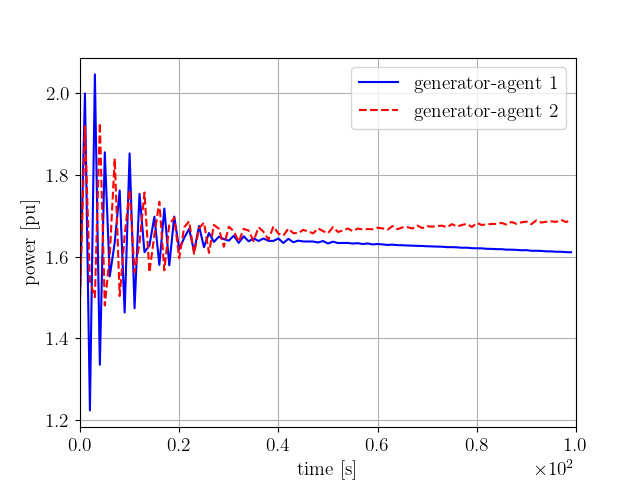}
                \caption{Generators' output in the first 100 s.}
                \label{fig16}
        \end{subfigure} 
        \hspace{2pt}
          \begin{subfigure}[b]{0.42\textwidth}
                \centering
                \includegraphics[width=\textwidth]{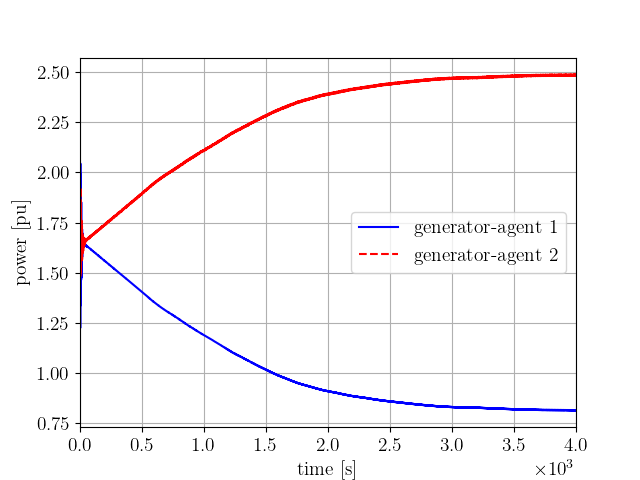}
                \caption{Generators' output.}
                 \label{fig16b}
        \end{subfigure}  \\
        \begin{subfigure}[b]{0.42\textwidth}
                \centering
                \includegraphics[width=\textwidth]{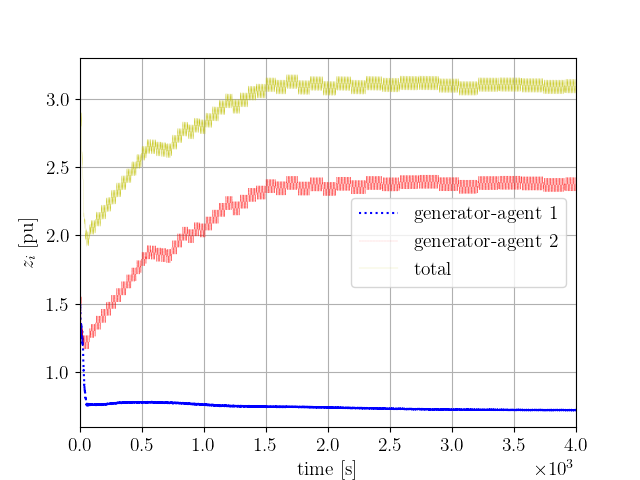}
                \caption{Secondary control action.}
                 \label{fig17}
        \end{subfigure}
        \caption{Secondary Control Model II: Change in load by $0.15$pu.}
          \label{figsub2}        
\end{figure*}

The reward function has been derived following \eqref{eqr_primary2}. Setting $\epsilon = 0.05$ pu; $d_1' = 100$; and $d_2' = 200$. The reward function is formulated as follows:
\begin{equation*}
r = \begin{cases}
100,              & \exists i: |\Delta \omega_i|<0.05 \\
200,&|\Delta \omega_1|\land|\Delta \omega_2| < 0.05 \\
0, & \text{otherwise}
\end{cases}.
\label{eqr1}
\end{equation*}

In Fig.~\ref{fig14} the cumulative reward obtained by the agents during training can be seen. Again, we notice that the agents are learning and have discovered how to obtain higher rewards. In this case, the agents learn how to jointly balance generation and demand.

Following the same schema, we change both loads by $0.15$ pu and then we observe how the frequency and each generator output change. The rotor electrical angular velocity of each generator is restored as it may be seen in Fig.~\ref{fig15}. The generation output of the two generators may be seen in Figs~\ref{fig16} and~\ref{fig16b}; Fig.~\ref{fig16} is a zoomed in version of Fig.~\ref{fig16b}. In Fig.~\ref{fig16}, where the timescale is up to 100 s, we notice that the total power of the generators meets the new load; thus restoring frequency. However, the secondary control system sends signals to the generators to modify their output as seen in Figs~\ref{fig16b} and~\ref{fig17}. The system frequency is nominal since even if the output of the two generators changes the summation of the output remains constant and equal to the new load.

We demonstrated that the proposed framework may be applied to solve primary and secondary control problems with the detailed modelling with Model II. This is achieved in a distributed way, i.e., without centralising any kind of information, the agents learn how to balance the system. Here, the agents learn that keeping $\Delta \omega_i$ close to 0 for all generators is associated with high rewards.

\subsection{Tertiary Control -- Model I}

The test case designed to test the performance of all levels of load frequency control in a single BA area comprises of two generation units or agents that interact with a load whose configuration during training may be found in Fig.~\ref{fig7}. The parameters of the environment can be found in Table~\ref{tab1} with cost functions for generator 1 $c_1 = 2P_1^2$ [\pounds/pu] and generator 2 $c_2 = P_2^2$ [\pounds/pu]. In each  episode, or training simulation, the load varies randomly around a nominal set point. The load varies as $P_L\pm\Delta P_L = 3 \pm \beta$ pu, where $\beta$ follows a uniform distribution.

The reward function has been derived following \eqref{eqr}. We set $\epsilon_1 = 0.05$ pu, $\epsilon_2=0.2$ pu, $d_1 = 200$, and $d_2 = 100$. Thus we have the two conditions:
\begin{equation*}
C1: |\Delta \omega |< 0.05,
\end{equation*}
\noindent and
\begin{equation*}
C2: |2z_1 - z_2|< 0.2.
\end{equation*}
\noindent Taking these two conditions into account we may formulate the reward function as
\begin{equation}
r = \begin{cases}
   200,& \text{if } C1 \land C2 \\
    100,              & \text{if } C1 \lor C2 \\
    0, & \text{otherwise}
\end{cases}.
\label{eqr1}
\end{equation}

\begin{figure*}[t!]
\vspace{-1\baselineskip}
        \centering
        \begin{subfigure}[b]{0.42\textwidth}
                \centering
                \includegraphics[width=\textwidth]{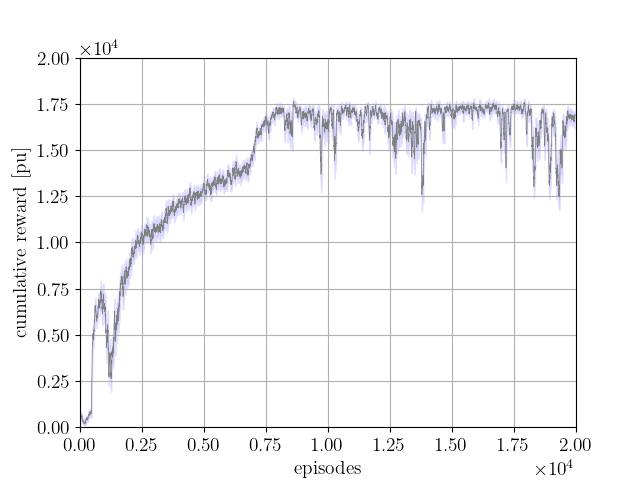}
                \caption{Smoothed cumulative reward per episode with 95\% confidence levels.}
                \label{fig8}
        \end{subfigure} 
        \hspace{2pt}
          \begin{subfigure}[b]{0.42\textwidth}
                \centering
                \includegraphics[width=\textwidth]{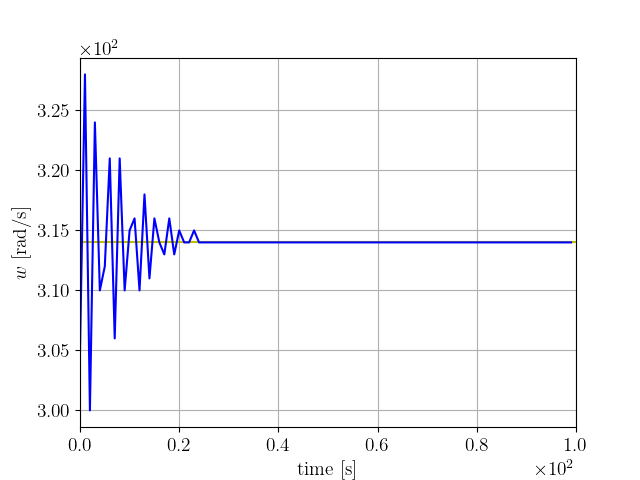}
                \caption{Centre of inertia speed.}
                 \label{fig9}
        \end{subfigure} \\
        \begin{subfigure}[b]{0.42\textwidth}
                \centering
                \includegraphics[width=\textwidth]{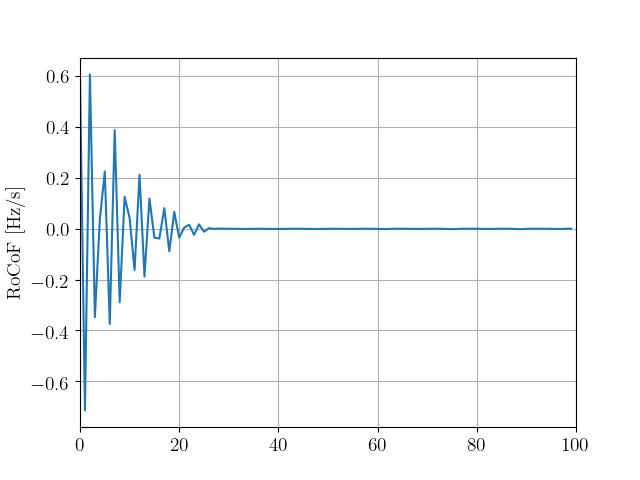}
                \caption{RoCoF.}
                \label{fig_rocof_normal}
        \end{subfigure} 
        \hspace{2pt}
          \begin{subfigure}[b]{0.42\textwidth}
                \centering
                \includegraphics[width=\textwidth]{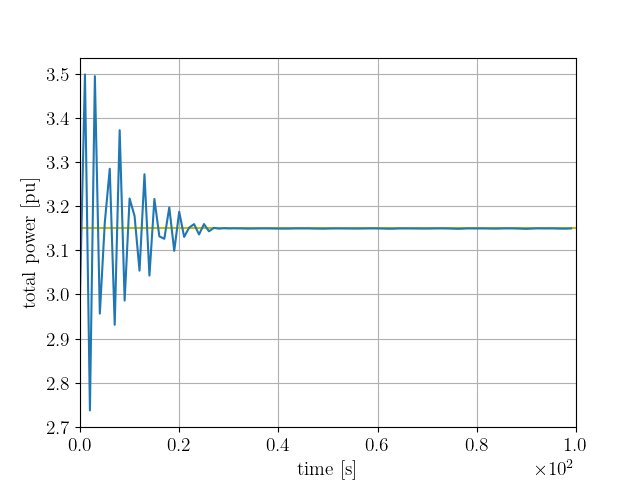}
                \caption{Total power.}
                 \label{fig10}
        \end{subfigure}  \\
        \begin{subfigure}[b]{0.42\textwidth}
                \centering
                \includegraphics[width=\textwidth]{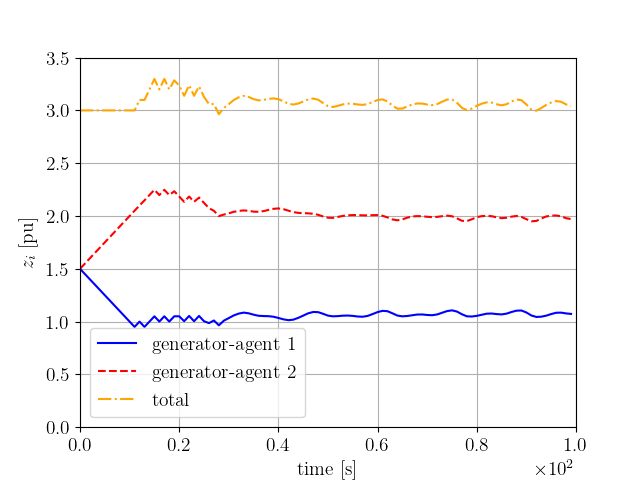}
                \caption{Secondary control action.}
                 \label{fig11}
        \end{subfigure}
        \hspace{2pt}
          \begin{subfigure}[b]{0.42\textwidth}
                \centering
                \includegraphics[width=\textwidth]{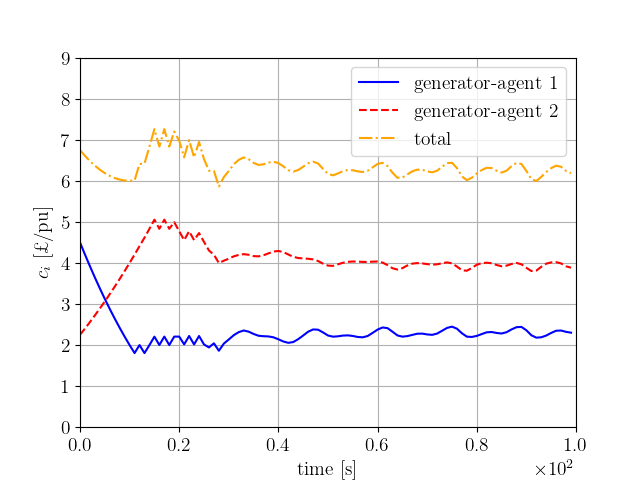}
                \caption{Cost of the generators.}
                 \label{fig12}
        \end{subfigure}
        \caption{Tertiary Control Model I: Change in load by $0.15$pu.}
          \label{figsub3}        
\end{figure*}

The reward function is used only during the training period. In the operation phase, the actors interact with the environment without experiencing any reward. Agents only observe the frequency of the system and their own control action $z_i$. They have learnt during training how to behave according to the evolution of the environment to balance supply and demand while minimising operational costs. For the operation phase, we change the load by $0.15$ pu and then, we observe how the agents restore the system frequency.

We can observe in Fig.~\ref{fig8} the cumulative reward obtained by the agents. The agents can obtain $20,000$ at maximum per episode, i.e., the maximum reward per step is $200$ and the number of steps per episode is $100$. The agents learn how to obtain higher rewards as the number of episodes increases. If that was not the case the cumulative reward function would oscillate around small values near zero.

In Figs~\ref{fig9},~\ref{fig10} we see how the agents restore the frequency to the nominal set point, thus balancing supply and demand. { \color{black}In Fig~\ref{fig_rocof_normal} the rate of change in frequency (RoCoF) that measures the dynamic performance of the system is depicted. The maximum, minimum and mean RoCoF values are $0.607$, $-0.712$ and $0.002$ respectively, thus being within the admissible limits of $1$ Hz/s recommended by ENTSO-E ~\cite{entsoe}.}  Actors learn how to balance generation and demand without exchanging information. The agents have learnt that keeping $\Delta \omega $ close to 0 is  the key to obtain high rewards. Thus, the agents may perform primary and secondary control in a totally distributed manner.

In order to test the optimality of the solution provided by the proposed approach  in terms of cost, we need to calculate the optimal point when the load in the system is $P_L =3+\Delta P_L = 3.15$ pu for the cost functions given in Table~\ref{tab1}. By solving the economic dispatch problem as given in~\eqref{eq:ed} we have $P_1 = 1.05$ pu and $P_2 = 2.10$ pu. In Figs~\ref{fig11},~\ref{fig12}, the behaviour of each generator output and its associated cost are depicted. It can be observed that the agents operate near the optimal solution, which is that the generation of generator 2 is twice that of generator 1.  As seen in Fig.~\ref{fig11}, the control action of agent 1 stabilises around a set point that is approximately half of the control action of agent 2. This does not coincide with the optimal solution (slightly above half the production, i.e., 60\%), but through the training process the agents learn how to keep load and supply balanced in a fully distributed cost efficient way. The performance of the agents is determined by what actions they learn during training that lead to high rewards. Thus, the reward function is the main tool to show each agent what is the optimal action. The reward function defined in~\eqref{eqr1} builds a reward combining two different dimensions: cost and frequency. This means that the reward function can show various maxima depending on the combination of both reward dimensions. The agents learn by trial and error a behavioural heuristic to obtain high rewards, but they may converge to a local optimum that may be different from the global one. An improvement of the reward function~\eqref{eqr1} could help the agents to improve the results showed here and to get closer to the optimal solution.

{ \color{black}

We compare the proposed framework with ~\cite{7163587}, neglecting the network effects so that the comparison is fair since in Model I we do not consider the network. In~\cite{7163587}, a distributed load frequency control algorithm that restores system frequency in a cost effective way is presented. This is achieved by exchanging some information between the generators during the operation phase. The algorithm is based on a partial primal-dual gradient scheme to solve the optimal load frequency control problem, this type of dual-based approach is fairly common in the literature of the minimum cost load frequency control problem. We refer to this algorithm as benchmark algorithm. In Figs.~\ref{fig_w_dist},~\ref{fig_p_dist} it can be seen that the benchmark algorithm manages to balance generation and demand, although it converges slightly slower than the proposed approach. In Figs.~\ref{fig_z_dist},~\ref{fig_c_dist} the secondary control action and cost of each agent are shown. The response of the benchmark algorithm is smoother than the proposed approach and the generation cost is minimised. However, this solution still needs to share dual information across units. On the other hand, although there are no optimality guarantees in the proposed framework, the results show that a sub-optimal solution is reached, it is fully distributed, i.e., no information is shared between the agents, and they only use local information.

\begin{figure*}[t!]
\vspace{-1\baselineskip}
        \centering
          \begin{subfigure}[b]{0.42\textwidth}
                \centering
                \includegraphics[width=\textwidth]{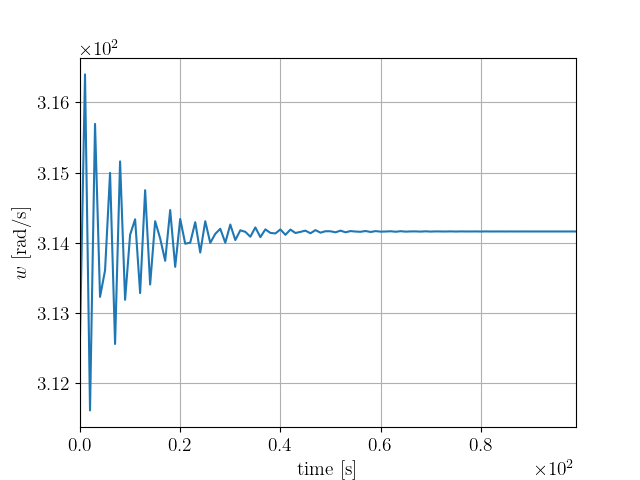}
                \caption{Centre of inertia speed.}
                 \label{fig_w_dist}
        \end{subfigure}   
        \hspace{2pt}
        \begin{subfigure}[b]{0.42\textwidth}
                \centering
                \includegraphics[width=\textwidth]{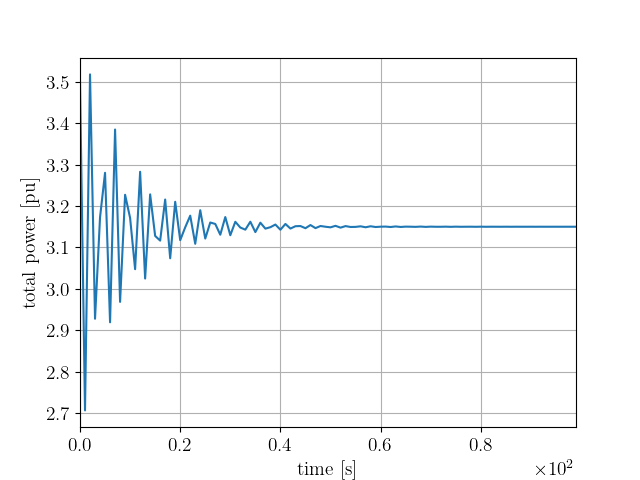}
                \caption{Total power.}
                \label{fig_p_dist}
        \end{subfigure} \\
          \begin{subfigure}[b]{0.42\textwidth}
                \centering
                \includegraphics[width=\textwidth]{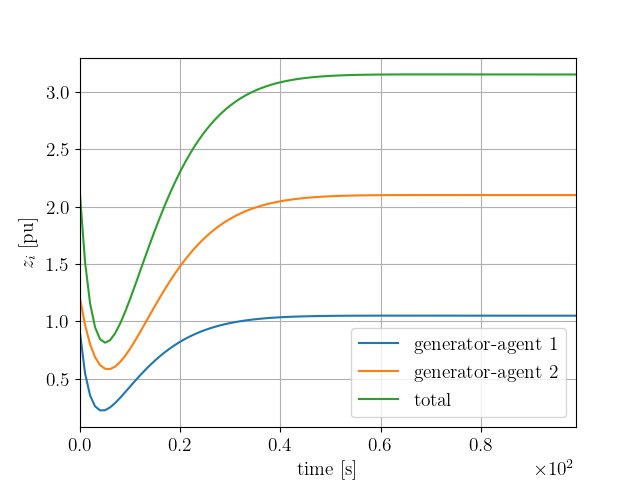}
                \caption{Secondary control.}
                 \label{fig_z_dist}
        \end{subfigure}  
         \hspace{2pt}
        \begin{subfigure}[b]{0.42\textwidth}
                \centering
                \includegraphics[width=\textwidth]{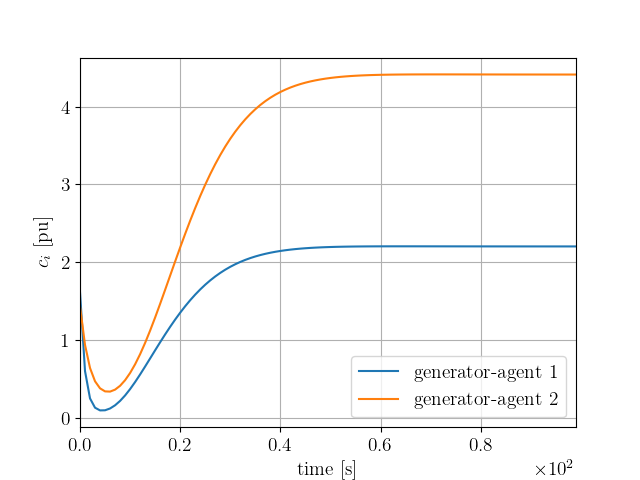}
                \caption{Cost of the generators.}
                 \label{fig_c_dist}
        \end{subfigure}
        \caption{Change in load by $0.15$pu using the benchmark algorithm.}
          \label{figsub4}        
\end{figure*}

\begin{figure*}[t!]
\vspace{-1\baselineskip}
        \centering
          \begin{subfigure}[b]{0.42\textwidth}
                \centering
                \includegraphics[width=\textwidth]{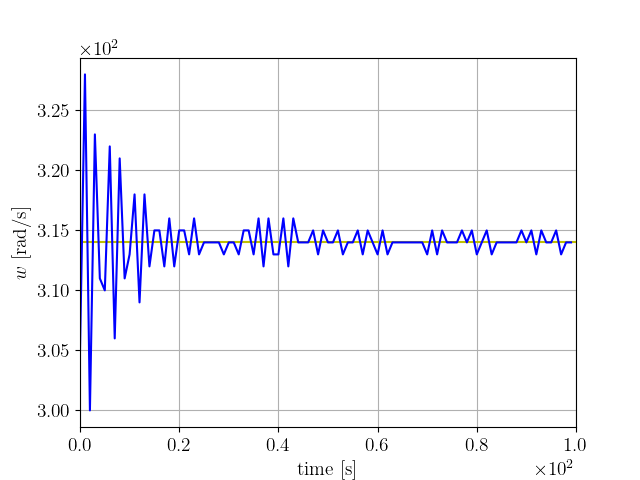}
                \caption{Centre of inertia speed.}
                 \label{fig_w_dyna}
        \end{subfigure}   
        \hspace{2pt}
        \begin{subfigure}[b]{0.42\textwidth}
                \centering
                \includegraphics[width=\textwidth]{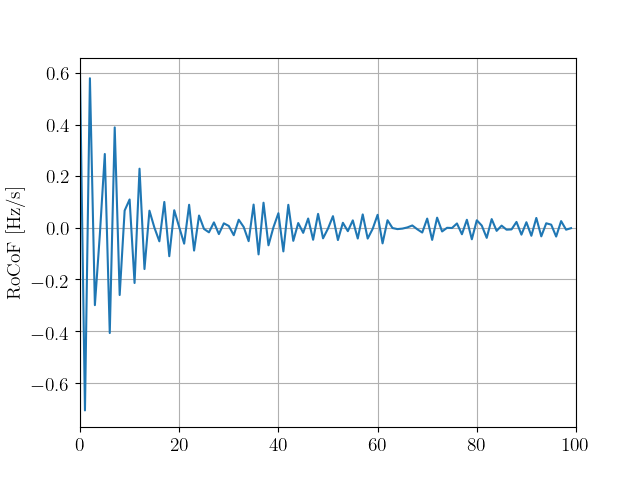}
                \caption{RoCoF.}
                \label{fig_rocof_dyna}
        \end{subfigure} \\
          \begin{subfigure}[b]{0.42\textwidth}
                \centering
                \includegraphics[width=\textwidth]{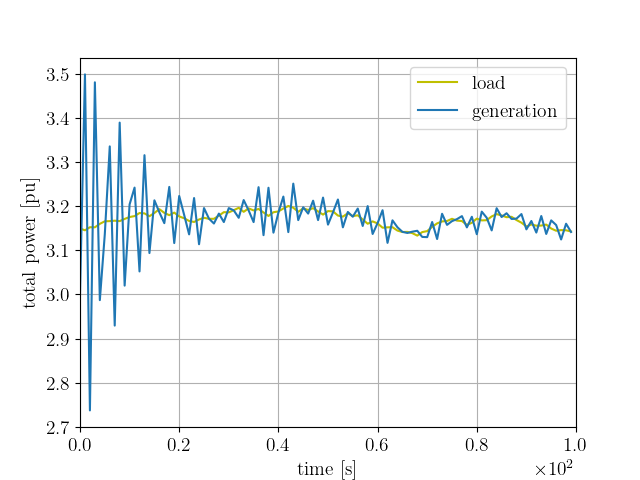}
                \caption{Total power.}
                 \label{fig_p_dyna}
        \end{subfigure}  
         \hspace{2pt}
        \begin{subfigure}[b]{0.42\textwidth}
                \centering
                \includegraphics[width=\textwidth]{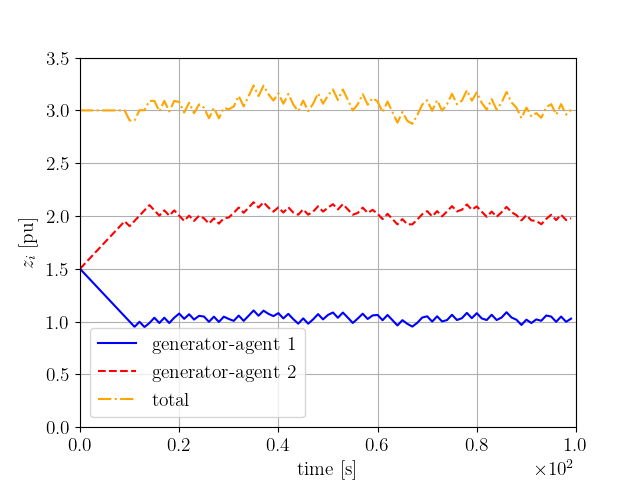}
                \caption{Secondary control.}
                 \label{fig_z_dyna}
        \end{subfigure}\\
        \begin{subfigure}[b]{0.42\textwidth}
                \centering
                \includegraphics[width=\textwidth]{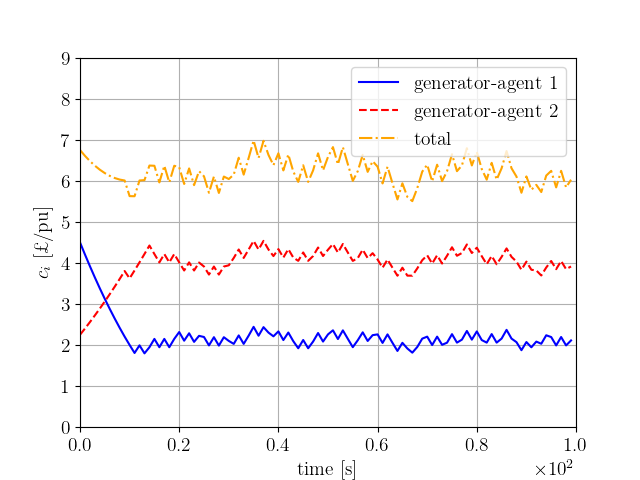}
                \caption{Cost of the generators.}
                 \label{fig_c_dyna}
        \end{subfigure}
        \caption{Tertiary Control Model I: Change in load by $0.15$pu followed by continuous changes in the load.}
          \label{figsub5}        
\end{figure*}

We also run a numerical experiment implementing a more realistic scenario, where an initial load increase of $0.15$ pu is followed by a continuous change in the load sampled from a uniform distribution defined in the $[-0.1, 0.1]$ interval. We can observe in Figs.~\ref{fig_w_dyna},~\ref{fig_p_dyna} that the agents manage themselves to keep generation and demand balanced, although the load is continuously changing. The dynamic behaviour of the system is depicted in Fig.~\ref{fig_rocof_dyna}, where it can be seen that the RoCoF does not go beyond of the admissible $1$ Hz/s bound (maximum, minimum and mean RoCoF $0.595$, $-0.708$ and $0.002$ respectively). Interestingly, it is shown in Figs.~\ref{fig_z_dyna},~\ref{fig_c_dyna} that the agents keep generating in a close-to-optimal ratio despite the continuous change in the load that increases the difficulty of this task.

}

In the numerical studies we showed that load frequency control may be performed efficiently in a distributed manner. More specifically, we showed that instead of solving the economic dispatch to obtain the optimal operating point, the MARL framework may be used to infer the production costs and the necessity of balancing demand and supply from the reward function and enclose this information in the behaviour of the actors. The benefit of the proposed approach is that these agents can act in real time in a distributed way. Once trained, they do not need to centralise information at all. Dynamics are embedded in the agents that only use local information.

\section{Concluding Remarks}
\label{sec6}

In this paper, we proposed an MARL alternative to implement load frequency control in a distributed cost efficient way. To this end, we expressed the load frequency control problem in an MARL setup and designed the reward functions based on insights on the economic dispatch problem. We chose an appropriate algorithm, i.e., MADDPG, to solve this problem. Through the numerical examples, we showed that the proposed framework performs load frequency control in a satisfactory way. In particular, we demonstrated that all levels of control are 
achieved using Model I, i.e., restored frequency to the nominal value in a cost efficient way; and that secondary control is performed under detailed modelling of Model II. {\color{black} Moreover, we showed that the proposed methodology can cope with generation rate constrains and uncertainty sources efficiently.}

There are natural extensions of the work presented here. For instance, different elements of the MARL paradigm can be enhanced, i.e., the reward function, the LSTM architecture and the introduction of domain knowledge could be further analysed to come up with agents that are able to improve their performance. {\color{black} More specifically, other architectures such as gated recurrent units (GRU) could be used instead of a LSTM. An exhaustive search of the appropriate architecture, parameters and hyper-parameters are left to be explored in future work. Another obvious extension consists of adding the tertiary control layer to the network model.} In our future studies, we also plan on studying the applicability and scalability of these techniques in more complex scenarios. In addition, we will investigate the performance of MADDPG to deal with different types of generation resources. We will report on these developments in future papers.

\bibliographystyle{IEEEtran}
\bibliography{journals-full,references}

\balance

\end{document}